# ‘An evaluation of Flickr’s distributed classification system, from the perspective of its members, and as an image retrieval tool in comparison with a controlled vocabulary’

A dissertation submitted for an MA in:

**Information Services Management**

**London Metropolitan University**

**September 2008**

**By Samuel Piker**

## Abstract

The profusion of digital images made available online presents a new challenge for image indexing. Images have always been problematic to describe and catalogue due to lack of inherent textual data and ambiguity of meaning. Because professionally-applied metadata is not practical for most open, web-based collections a solution has been sought in the form of tags, simple keywords that can be attached to an image by any web user. Together tags form a flat structure known as distributed classification, or more popularly as a folksonomy.

This research follows the debate surrounding folksonomies and aims to fill the gaps in understanding of why people tag and how effective they find them for searching, using the photo-sharing website Flickr as the focus. Open-ended questionnaires were sent out to members of the site who use tags, with the opportunity to post comments to an online discussion space. The other key gap identified in the literature is a systematic comparison between a tag-based system and a more traditional controlled vocabulary, to test out the claims made regarding tagging’s suitability for searching and browsing. For this purpose Flickr has been compared with Getty Images using a series of test themes.

The small number of people who replied to the questionnaire gave detailed answers that confirmed several of the assertions made about tags: they are accepted despite their flaws (sloppiness and potential for inaccuracy) because they serve their purpose to a satisfactory level. Some answers challenged the assumption that tagging is only done for personal benefit. The search comparison found that while Getty allows highly specific queries and logical semantic links, Flickr is more flexible and better placed to deal with subtle concepts. The overall conclusion is that tags achieve most when used in conjunction with groupings of people with a shared interest.

# Contents

## Statement of originality

This research is the work of Samuel Piker and has been completed solely in fulfilment of the dissertation for the MA in Information Services Management at the London Metropolitan University.

## Acknowledgements

Sue Batley, for sound advice, guidance, and ongoing feedback in her role as supervisor for this research.

The following colleagues of the author:
Learning Resources Manager Ann Badhams, for allowing study leave.
Faculty librarian Ursula Crow, for passing on relevant literature.
Bibliographic Services Librarian Mary Kirkness, for diligent proof reading.

The author's family, for ongoing encouragement and proof reading.

Assorted members of Flickr, for taking the time and effort to help with the pilot and answer questionnaires.

# 1. Introduction

The idea for this research arose from an article in a professional journal which coincided with a personal interest. Matusiak (2006) examined the application of user-created metadata to digital image collections, contrasting some examples from the Flickr website and a University image collection in the USA. This caught the attention of the author, who had been displaying photographs on Flickr for several months and who was also developing an interest in the field of professional indexing.

The Matusiak article drew some conclusions about the varying quality of the user-created metadata, that its quality, consistency, and depth tended to vary, whilst recognising the new opportunities it offered for representing users' mental models more closely than traditional forms of image indexing (p. 283). These conclusions appeared to be justified in the light of the examples that were examined: individual images along a theme, with accompanying metadata. Yet the complexity of the subject and its far-reaching implications for how online digital images are organised and retrieved called for a more complete body of evidence, something beyond a comparison involving four examples, to gain a deeper insight into the phenomenon.

During the summer of 2007 two articles written by Jonathan Eastland appeared in the British Journal of Photography, addressing the parallel developments of tagging in image collections, and the new micro-stock sites selling a large volume of generic images. Coverage of the evolution in image collections in the photographic community's main UK publication further reinforced the sense that research in this area would be timely and relevant.

## 1.1 The overall aim

If the growing phenomenon of a system's users assigning their own metadata to web-based digital images was to be addressed, then a collection that used such a method needed to be evaluated in detail. The author was already taking tentative steps onto Flickr (as a showcase for photographs) and the site had a high profile in the literature of the information profession. It made a logical focus for the research.

It is possible to break the process of image indexing into three key areas: description, organisation, and retrieval. As images carry no form of inherent textual data the only words that appear in the associated metadata are those assigned to them by people making conscious decisions (Enser, 2000, p. 200), so they have to be described in a way that indicates what they represent. As the reading stage would discover, such description can reach to a number of levels.

Digital collections are organised, consciously or not, so that any one item is related to others that share similar properties. As in the description, images can be linked to each other on different levels such as objects they portray, location, mood, or style (Rasmussen, 1997), and the more detailed the description then the greater capacity for creating associations. Description leads into organisation. What a controlled vocabulary does is to define the words that can be used to describe, and then make meaningful connections which form a layered structure of subjects. This has been the basis for traditional libraries and databases, but due to time and expense issues such a system is not always feasible in the digital environment, which is why other solutions are emerging.

The object of describing and organising images is to make them retrievable at a later stage (Chowdhury, 1999, p. 2). The ability to answer a person's request for an item is what makes an effective retrieval tool, whether this is achieved by accurate search results or metadata that leads them from one item to related ones. It is therefore possible to evaluate the effectiveness of an image collection in these three key areas, to base the measure of its success on how it allows images to be described, organised and retrieved by its users. This realisation helped to form the main aim of the research:

*Evaluate the success of Flickr's tagging-based system in describing, organising, and retrieving web-based digital images, in comparison with Getty Images.*

## 1.2 The focus of the research

The website Flickr started in 2004, offering a personal space to store digital images (primarily photographs, although scanned or computer-generated drawings and paintings do feature as well) and share them with other members, although it was also searchable by any web users. If members wished to they could make material available only for family or friends, and Creative Commons licensing was made available as an option with its varying levels of conditions on how images are used.

Tags are at the core of Flickr's system. They are simple keywords that members can choose to apply to their images, which then act as links to other images tagged in the same way. This can operate just within their own personal space or 'photostream', or across the whole of the collection. An individual can group their images into sets as well as adding them to the 'pools' of particular groups, alliances of members who have an interest in a certain subject. These features, and tagging in particular, can be referred to as distributed classification, in that it is not controlled and has no central structure.

## 1.3 Forming the objectives

One key feature of distributed classification is that the people who use the system to retrieve images are also the ones describing and organising the material. In this sense Flickr is a form of community, and the popular term for this type of system is a folksonomy (Vander Wal, 2005), a play on the word taxonomy, meaning a layered classification system, and a reflection of the open communal (folk) aspect.

Any evaluation of Flickr's effectiveness depends on gathering the opinions of its inhabitants. For this reason the first two objectives were as follows:

1) *Identify the most positive and negative aspects of tagging as an image description and organisation tool, from a tagger's perspective.*

2) *Evaluate the success of tags in retrieving online images, from the perspective of Flickr members.*

During the literature search for a related piece of academic work, a project proposal focusing on specialised image collections, a gap in the research quickly became apparent; the experience of those who populate sites like Flickr was an area waiting to be properly addressed. Part of the effectiveness evaluation was to address what it was like to describe and organise images, and tags are central to this. Therefore the first objective was to identify the members who use tags in order to find out what they regard as the most negative and positive aspects of the system.

As stated previously, the people that create tags also use them to look at other photographers' work on the site. In a continuation of the idea that Flickr's inhabitants hold the key to evaluating its success as a descriptive and organising tool, their experiences could also help judge its effectiveness in retrieving images. The thinking behind the second objective was to identify what the site was like to use as a retrieval tool for those who helped classify the material.

Looking at an issue from more than one angle helps to give a more complete picture, so while the views of Flickr members formed a large part of the evaluative exercise it was also important to take a methodical, critical approach. Because distributed classification is in such stark contrast to traditional indexing methods the two are inevitably compared by commentators such as Matusiak, but the exploratory literature search found no evidence of a systematic comparison. The third objective was designed to fulfil this aim:

3) *Analyse the effectiveness of Flickr's folksonomy as a web-based image retrieval tool, in comparison with a controlled vocabulary.*

Getty Images is covered in more detail in the methodology section, but its key difference from Flickr is that it uses a controlled vocabulary for the descriptive keywords assigned to each image. The fact that it is also a web-based image collection made it a logical point of comparison, with the focus on the retrieval side of the process.

# 2. Literature review

## 2.1 Introduction

Rapid developments in the technology of digital photography have coincided with advances in web access to dramatically increase the number of images available online. Our approach to organising information has begun to struggle to keep up with our ability to produce and store it, placing the majority of resources beyond the practical limits of traditional indexing. Images have always presented their own challenges in terms of the effort required to classify them, and without inherent textual data they require humanly assigned metadata for retrieval.

An apparent solution arose in the form of tags, short descriptive terms which could be added to any item, including digital images, which then acted as keywords for searching or hyperlinks to other items with the same tag. Together these tags form what is popularly known as a folksonomy. One of the most widely used services to employ such a system is Flickr, a site for storing and sharing personal photograph collections.

This review will track the key features of discussion in tagging's native web environment, in the popular press, and amongst the photography community. Information professionals have been quick to engage with the new approach to classification, although an examination of existing research will reveal key gaps in the understanding of the cognitive, emotional mechanism of tagging. Relevant issues of traditional image indexing will be addressed, as well as how elements of cultural criticism can help inform our understanding of tags and those who create them.

## 2.2 Commentary from the web logs (blogs)

The evocative term folksonomy was coined by web developer and information architect Thomas Vander Wal to denote the open assigning of descriptive terms to online objects, in a process known as 'tagging'. Also referred to as 'free tagging', 'ethnoclassification', and 'faceted hierarchy', folksonomy is defined as 'a user-based approach to organizing information that consists of collaboratively generated, open-ended labels created by users to categorise web information resources' (Chowdhury & Chowdhury, 2007, p. 108). These resources could be web pages themselves or individual items of media such as digital images.

It is appropriate to open the debate on this uniquely web-birthed phenomenon with a look at some key blog entries on the subject. Merholz advocates ditching the term folksonomy altogether due to its misleading implications of hierarchy, arguing that 'Tagging does not a taxonomy make' (2004). He views tags as being of most use to the taggers themselves, while any benefit to the wider community is merely a by-product. This need not be a negative thing, as residual benefit is no less valuable, and dismissing tags as being an 'utter disaster' for targeted searching seems a little exaggerated, although Merholz does also regard them as being ideal for serendipitous browsing. It will be interesting to see whether these views are shared by users.

In the light of such discussion Vander Wal has felt the need to clarify his invented term, asserting that while tagging is usually done for personal retrieval in a social context it is not necessarily collaborative by design (2005). As regards the lack of hierarchy he points out that 'The people are not so much categorizing as providing a means to connect items and to provide their meaning in line with their own understanding' (Vander Wal, 2005). While this does not exactly address the inaccurate inference of layered taxonomies, the point is that people are empowered by being able to express and organise resources in their own way.

In the same posting Wander Wal raises the idea that the real value of a folksonomy is in connecting people with similar interests (i.e. who tag similar objects in the same way). This is particularly relevant to a primarily amateur photo-sharing tool such as Flickr, and emerging research (Cox, Clough, and Marlow, 2008) suggests that the ability to share interests is one of the site's most valued aspects.

One side-effect of tagging's spread is that traditional indexing has started to appear more often in the discourse of non-information professionals, partly for purposes of comparison, but also in order to provide a frame of reference. Shirky views the growth of folksonomies as not only inevitable (2005a) but also more appropriate to a web environment. He regards tag-based systems as being part of a more organic approach to organising information, able to deal with a lack of stable entities or formal categories, and to accommodate amateur and uncoordinated users (2005b). Mejias (2005) compares the two approaches in terms of the categorization process (or lack thereof) and consensus; in a taxonomy the structure is agreed upon between indexers, whereas in tagging the consensus is assumed by the system itself at the moment a tag is created, so the need to negotiate meaning at the initial stage is simply bypassed.

This is not to say that tagging has an obscuring or isolating effect. Mejias and Shirky both consider that folksonomies 'seem to work rather well' despite the lack of structure (Mejias, 2005). In addition the quick access they offer to all the other taggers who have used an identical tag gives a very tight feedback loop, allowing users to gauge their tag's relevance by comparison, and most systems allow easy correction of tags across different items (Udell, 2004). While the signs are very promising, the assertions of these commentators demand to be tested against the experiences of those who inhabit tag-based systems.

Where Mejias differs is in his appreciation of how traditional indexing principles could enhance a folksonomy (2005), and the value of understanding people's interaction with a tag-based system. His small-scale study based on users of del.icio.us found that they tended to separate tags intended for themselves and those for a wider audience, indicating an awareness of communal benefit alongside serving their own information needs (Mejias, 2004). There was also a split between those who craved a bit more structure, and those who recognised the potential for multiple, complex subjects. It will be interesting to see whether similar usage patterns emerge in the case of Flickr's photographers.

## 2.3 Newspaper coverage

The lively discussion amongst bloggers and the growth of tagging services has generated a healthy amount of newspaper coverage. Articles in the media, education, and business sections of the Guardian and Observer have tended to focus on the potential for harnessing 'collective intelligence' (O'Hear, 2005) and the power of shared interests (Hind, 2005). Burkeman and Naughton (2005) both view folksonomies as resonating with users, and consider tagging the only feasible solution to organising massed online content, despite its lack of rigour.

Lilley offers a word of caution; even online communities are diverse in makeup, and viewing user-generated content as a single entity is lazy thinking (2006). Indeed, there is a disparity between Burkeman's claim that folksonomies threaten professional indexers and his admittance that they 'aren't much good for detailed research'. Most notably there is lack of awareness amongst press commentators about how tagging systems themselves can be improved upon.

## 2.4 Photographers' press coverage

Another group taking interest in tag-based systems are the professional photographers, who have a new opportunity for selling high-quality, generic images on the various 'micro-stock' sites. As Eastland of the British Journal of Photography points out, 'if you want your work to be seen in this global showcase, they need to be properly and comprehensively described' (2007b, p. 23). Eastland goes on to point out that even Flickr's accessible tagging system is still under-used, and that synonyms do not constitute semantic links (p. 24). If the former is true, then contact with Flickr's inhabitants should give some clue as to why this is the case.

It is revealing that the micro-stock site iStockphoto has moved from free tagging to a controlled vocabulary to which all tags have to be mapped, as well as issuing a set of guidelines to their contributors. Their argument for combating 'tag spam' is that 'Using irrelevant tags is detrimental to everyone in the community' (Lane, 2007), which seems to support Mejias's findings regarding people's motivation for creating good metadata.

## 2.5 Traditional image-indexing issues

Creating metadata for images has always been a problematic process. Eakins refers to manual indexing as 'inherently both unreliable and labour-intensive' (1998, p. 5), the latter problem stemming from the fact that descriptive keywords have to be created as a 'textual surrogate' of the image (Enser, 2000, p. 200). The 'intellectual and practical challenges posed by the semantic indexing of images' (Enser, p. 201) occur particularly at the abstract level, trying to describe types of activity or higher concepts (Eakins, p. 3). These issues are not unique to images but they are accentuated in the absence of any inherent textual data.

Because of the abstract nature of much image metadata, assigning it tends to be a highly subjective process (Enser, p. 201), and due to the cultural nature of interpreting images 'Interindexer consistency has always been problematic' (Chen and Rasmussen, 1999, p. 294). If this is the case with trained indexers, putting metadata creation in the hands of web users is not as great a compromise as it might first appear. It also follows that amateur taggers will face the same challenges of analysing meaning.

One tool which can help reduce uncertainty is a controlled vocabulary, as mentioned earlier in the case of iStockphoto. As this site has recently been bought by Getty Images it is likely that it uses the same Art and Architecture Thesaurus (AAT), one of the most widely used tools for indexing images. While Rasmussen refers to the AAT as 'a carefully constructed vocabulary' (1997, p. 179) she goes on to warn that it lacks coverage at the cultural, conceptual level (Chen and Rasmussen, p. 295). With this in mind, a search comparison between Flickr's open tagging and Getty Images could reveal certain advantages in the looser approach.

An effective retrieval system does not only deliver precise results to a query but also allows the user to browse. Foskett is a firm advocate of serendipitous discovery, saying 'it is often an item which does *not* fit our existing patterns of interest which proves to be the most interesting' (1996, p. 26). Enser also advocates this approach, and he believes that because semantic indexing is so problematic, 'browsing usually plays a highly significant role in visual information retrieval' (p. 206). In any assessment of a system's effectiveness it is therefore important to include its browsing facilities.

## 2.6 Professional literature

If the rise of distributed classification has got the passing attention of the popular press it has generated an even greater degree of interest amongst information professionals; this includes those in the wider field of information science along with practitioners in librarianship. Much of the discussion has taken place in the online journals, although the measure of the profession's interest is best encapsulated by the inclusion of folksonomies in textbooks by Morville (2005) and Chowdhury and Chowdhury (2007). Much of the professional literature is driven by a desire to address the lack of real, empirical evidence emerging from the blogosphere (Sinclair and Cardew-Hall, 2008, p. 17), and the body of research is still in its infancy.

Exactly how to define the new phenomenon is a crucial point of debate. While the term 'folksonomy' is widely recognised as the dominant label, 'user-indexing', 'mob-indexing', 'folk classification', and 'social classification' have all been put forward in the interests of accuracy. Since the tagging process is specifically a type of classification rather than general indexing (notation is not usually involved and there is no prescribed record format), and it can be social but does not have to be, 'distributed classification' would seem to be the most appropriate. The other terms may of course be used where collaboration is the core element, or if there is some form of agreed format.

Certain patterns emerge from the discussion. Distributed classification schemes are considered to be unsophisticated by traditional library standards, but they 'make more sense to the user community' (Chowdhury & Chowdhury, p. 108), use the 'collective expertise', and 'empower the user' by being meaningful to them (Fox, 2006, p. 169). Chowdhury and Chowdhury focus on the potential of social classification to help bring order to the Web (p. 109), which Hammond et. al. (2005) regard as being 'still a teenager and subject to all the angst, vagaries, and contrariness of growing pains'.

Matusiak views social classification as offering at least a partial solution to the challenges of image indexing, in that it makes use of user expertise and communal verification (2006, p. 288). In her sample of Flickr images the quality and volume of tags varied widely, and they lacked the consistency and hierarchy of library metadata. The perceived advantage of Flickr's system is a greater capacity for multi-lingual tagging and the opportunity for users to 'describe the world in the way that they see it' (p. 294). It is worth noting that the former point does not account for the capacity of controlled vocabularies to translate automatically, and the latter is based largely on inference rather than contact with users themselves.

An idea shared by both Fox and Morville is that the close connection and fluidity of distributed classification could help enrich the more controlled vocabularies. In a process known as 'pace layering' (Morville, p. 141) the fast-adapting folksonomies would catch any emerging trends and collect feedback, while the more traditional schemes provide a more stable layer and respond at their own pace.

Enhancement can also go the other way. Guy and Tonkin (2006) have offered ways to improve tagging with traditional indexing principles, along the same lines as Mejias. Their recommendations for improving precision and encouraging socially beneficial tags include providing a checklist to help users draw out the 'salient characteristics' of an item, using plural forms, and using the underscore for compound_terms. They also appreciate the danger of tidying up too much and losing the essential fluidity which makes tagging systems so appealing to their users.

## 2.7 Emerging research

A key problem raised by Guy and Tonkin is the scarcity of qualitative data on the thought process behind tag selection and modification. Golder and Huberman's analysis of patterns in del.icio.us tags (2006) found that taggers tended to emulate each other, although minority opinions were still able to exist. They expect that similar patterns will be played out in other collaborative tagging systems.

Cox, Clough, and Marlow at the University of Sheffield have taken a more individual approach with Flickr, conducting 30-40 minute in-depth interviews with 11 users. Their focus was on how the site functions as an outlet for serious amateur photographers, although tags were touched upon. The majority of interviewees saw tags as important yet also boring and difficult. There was a roughly even split between those who regarded adding tags to others people's photos as intrusive, and the rest who considered it an act of public good to improve the accuracy of the metadata. Although laziness was admitted as a barrier to tagging, exactly why people found it challenging was not drawn out.

Another interesting discovery that the Sheffield researchers made was that Flickr users actually enjoyed the experience of 'getting lost', using the tags and comments to browse. As they put it, 'from the user's point of view the natural way to explore photographs is through people'. The browsing element has been explored by Sinclair and Cardew-Hall, who looked at the 'tag cloud' (a weighted list; frequency of use equals larger font) as a retrieval tool. A simulated dataset indicated that this feature was best suited to loose browsing and broad categorisation, while participants preferred a search interface for more specific queries (p. 27). A good system should allow fluid transition between the two modes.

## 2.8 Cultural criticism and tagging

A range of concepts have a bearing on the tagger's experience, including ideas of the self, language, and meaning. Some understanding of these areas can help make sense of what is taking place, and Day (2005) has proposed that poststructuralist criticism can be applied to new forms of communication and discourse in information studies. He is particularly interested in the way it calls the classical ideas of the 'user' and representation into question (pp. 601-602), and Weight believes that 'Digital textuality shares a capacity for variation with oral tradition' (2006, p. 435), the transience of language enhanced by cyberspace.

It is possible to view tagging as a sort of avant-garde metadata, moving away form established rules to forge new practices in image indexing. Lyotard considers 'Great joy is to be had in the endless invention of turns of phrase, of words and meanings' (1984, p. 10), and he talks of the search for instability in postmodern science. Furthermore he suggests that postmodern language games are about self-knowledge, allowing the community to reinvent its internal communication (p. 62). Therefore an analysis of distributed classification in its uneasy, unstable state would be improved in part by a postmodern perspective on tagging.

## 2.9 Conclusion

The consensus emerging from the blog entries and professional literature is that distributed classification is a better than nothing solution where expert indexing is not practical. Some commentators go further to suggest that it is actually more appropriate for the immediate, fluid nature of web resources, and it contains the potential to tackle the older challenges of image classification.

The split emerges between those who simply accept tagging in its current state, and those who regard this as an initial stage to be improved upon. The expressed desire for higher quality tagging amongst the photographic community suggests that taggers could be better informed and equipped, although making this happen without constraining them will be problematic. If tags appear to threaten professional image indexing, some comfort can be drawn from the postmodernist idea that if conventions are challenged or subverted, it is generally in order to put new ones in place or amend an existing system for the better.

Above all there is a need for greater understanding of the thought processes behind tagging, how well these systems work for people and what challenges taggers are facing. Early investigations suggest that browsing is central, although not in the sense of layered categories, and that sharing of tags works towards a shared descriptive language. Flickr also deserves to be analysed as an image retrieval tool in its own right, taking into account its powerful browsing facilities. The controlled vocabulary of Getty Images is ideal for such a purpose.

# 3. Methodology

## 3.1 Introduction

The following section traces the alterations made to the dissertation plan in response to the reading, and the specific strategies used during the literature review. These involved a combination of information science, newspaper, and communications studies databases, the reference lists of articles thereby located, and a search of university library shelves using previously acquired knowledge of key texts on organising information. One of the key realisations made is the extent to which humanly applied metadata depends on language and meaning. As a result of the questionnaire piloting the sample was changed to focus on those who tag their material in detail; the first two objectives were focused onto one questionnaire instead of two. Subjectivity of an image's meaning was acknowledged in designing the search comparison, which was modified slightly to test overall effectiveness in response to some complex and ambiguous queries.

## 3.2 The working title

The first change from the original dissertation plan was the working title. During the reading it emerged that evocative and catchy as the term folksonomy is, it is also problematic and misleading, so:

> *'An evaluation of Flickr's folksonomy from the perspective of its members, and as an image retrieval tool in comparison with a controlled vocabulary.'*

Became:

> *'An evaluation of Flickr's distributed classification system, from the perspective of its members, and as an image retrieval tool in comparison with a controlled vocabulary.'*

Distributed classification was settled on as more accurate, denoting a lack of hierarchy and central control. Likewise it needed to be made clear that the people on Flickr are part of an entity rather than just contributing to it. Evaluation remained a core element, although it was clearly necessary to explicitly separate the evaluation of the members' perspective from Flickr's performance in retrieving images.

## 3.3 The literature review

The original idea for this research grew out of the Matusiak article, read in preparation for an exam on indexing. The author also regularly browses the British Journal of Photography and the Eastland articles on tagging and micro-stock sites reinforced the felt need for an observational study of the online image-tagging culture.

The Emerald database was known to be a good source of scholarly articles from the information profession, whose perspective on an emerging form of indexing was considered important. A search turned up several articles on the broader subject of distributed classification, some of them touching upon the issue of digital images. The reference lists for these articles pointed towards a range of other relevant material, much of it web-based. A web search also uncovered the literature review of another library student (Speller, 2007) on the tagging of digital music, which itself provided many useful references. Due to the emerging nature of the subject e-journals and blog entries formed a large part of the reading, giving a very current overview of the debate and the key commentators.

The substantial amount of newspaper coverage was gathered together using the Lexis Nexis news database, where simple keywords such as 'flickr' and 'folksonomy' proved to be very effective in retrieving articles. This also proved to be the case elsewhere, although in the Emerald database technical terms like 'image indexing' and other descriptors from previously found articles were used to broaden results.

Such an important development in the area of indexing required a theoretical perspective to place the contemporary comment in some sort of context. The key texts in the organisation of information were selected mainly by previous studies of this area by the author, followed by a browse of the University's library shelves at the same location as these texts. Browsing was also used within the contents pages of any information journals that held an article that had been found by its reference elsewhere, as many of these journals have a common theme to a particular edition. Meanwhile a subject search of image indexing on the library catalogue produced a few key papers by Eakins; they were written before digital photography and the interactive web, but they helped to outline the main issues surrounding the metadata.

Something highlighted by many of the key texts on indexing is that classification and metadata is a form of language, a way of communicating ideas. This led to a search of Sage's Communication Studies database, which held several articles about communication in a digital context. Since the field of critical theory had produced an interesting angle on the subject, an already known core text by Lyotard was given another reading.

One other important source of relevant material was people; library colleagues who were aware of the dissertation topic pointed out articles they had come across on websites or in journals. Given the communal aspect of 'folksonomies' this seemed a very appropriate way to locate the information.

## 3.4 The questionnaire

The overall aim of the research was to evaluate the success of Flickr's tagging system, and one of the best indicators of this is the taggers themselves, those who upload and label their images. Building up a picture of what tagging is like and how it works was going to mean approaching Flickr members and drawing out their opinions.

In terms of access to the group being studied, Flickr has an internal mailing system for messages between members; the author is a member, so it made sense to use the available facility to contact people. One benefit of this was approaching taggers as a fellow member of a community, rather than a complete outsider.

The initial idea was to identify two sorts of Flickr members; those who used tagging extensively, and those who tagged very sparingly. The thinking was that this could help identify some of the negative aspects, or things that discouraged participation, as laid out in the first objective. The latter group would also be used to gauge the success of tags in retrieving images.

The chosen method was an open-ended questionnaire. This had the advantage of being distributable by Flickr mail, thus being no-cost and simple. In addition it fulfilled the remit of generating an in-depth picture of the taggers' experiences, in keeping with the observational, primarily qualitative nature of the study. The potential downside of this qualitative approach was how labour-intensive it could be; messages can only be sent on Flickr mail one at a time, so sending out enough to get a worthwhile response threatened to be time consuming, and in depth answers would require detailed analysis.

Despite the potential challenges, getting inside a tag-based system and asking open questions was deemed to be the best way of evaluating the tagger's experience. Distributed classification is social by its nature, so it made sense to incorporate some of Flickr's social features into the research style. To this end an image (a photograph of the hand-written words "Tagging questionnaire: further thoughts") was posted onto the author's page and its URL included at the end of the questionnaire, accompanied by an invitation to post comment underneath that could be seen and added to by others. The idea was to create an online discussion between the participants, hopefully generating even more interesting data.

### 3.5 Questionnaire design

In the original plan it was determined that the first questionnaire would be aimed at objective 1; identify the most positive and negative aspects of tagging. Meanwhile the second one would address the success of tags in helping retrieve other people's images, as outlined in objective 2. It was quickly realised however that the frequent taggers would have something interesting to say about the retrieval aspect, so questions on this were included in both versions.

The style of the questions was kept as brief and plain as possible; the best way to draw out someone's opinion on a subject in written form is to ask directly, and succinctly-worded questions encourage participation more than long-winded ones. Likewise the covering statement was worded to make the intentions and purpose of the research clear, an appeal for a moment of someone's time and thoughts.

### 3.6 Piloting the questionnaires

Before sending emails out it was important to test the questionnaire's effectiveness in generating interesting answers. The author has several friends and fellow students who are Flickr members, and amongst them there was a fairly distinct split between those who tagged in detail and those who added few tags if any to their work. This made an ideal pilot population.

The participants of the first questionnaire were the sort of Flickr users who add tags to any image they load onto the site. They selected tags using elements such as photograph genre, physical objects and their surroundings, a general subject, and abstract tags. There was also a comment about tags placed for comic effect that was not entirely clear. The lesson: one downside to the emailed questionnaire is that some unclear answers may have to be accepted without any further clarification. Forming questions as clearly as possible can help minimise this effect.

A supplementary question about what features were used (such as the 'choose from your tags' function) caused some confusion and was not of great importance, so it was cut out. This left the first question as 'How do you select your tags?', with 'What elements of a photo do you use to select them?' underneath in brackets to provide some extra clarity.

In answer to 'Why do you tag your photographs?' one person answered that it was mainly to search through their own photographs, while increasing the exposure of their work. To 'show off' and increase the posting of comments by others was the main motivation for another user. This user also made the following comment:

> Another reason is that Flickr is a resource: if I want to see a photo of Tower Bridge at night I know that by searching with the right tags I'll get ten thousand appropriate pictures!

(BUT I belong to the Guess Where London Group, where the point is NOT to tag photos until they're guessed.)

Considering this was part of the answer to why they tag their own photos, this implies an acknowledgement that tags help increase Flickr's performance as a whole, and also reveals that communal tagging can work as a fun activity.

The questions on whether tagging is enjoyable and how it could be made easier generated some interesting answers. The following sequence of questions however, turned out to be problematic:

Do you use tags to search and browse other people's photos?

If so, do the results reflect their tags accurately?

How useful do you find them in searching Flickr?

The one-word answers to the first one fulfilled its purpose, but there turned out to be a significant overlap between the second and third, so these were merged into simply 'If so, how accurate are the results of your search?'. In answer to how the quality of tags could be improved there was a suggestion for a 'fuzzy search' to handle spelling variations and hyphenations. This sort of creative suggestion boded well for the real research.

The answers to the non-tagger's questionnaire reflected a very different use of the Flickr site, mainly that of sharing images with close friends or fellow participants from an event. This should not have been a great surprise, and there were still some thoughtful answers, but it raised the issue of whether this second questionnaire was really necessary or relevant; the aim of the research was to investigate tagging, not why some Flickr members chose to opt out. As a result it was decided to drop the second questionnaire.

In the case of two participants in the pilot, the entire completed questionnaire was posted as a comment under the 'further thoughts' image. This was unexpected , but not necessarily a bad thing as it could help to get a discussion going.

## 3.7 Sampling strategy

The target of the research was people who tagged their work with some degree of detail. Beyond this criterion there were no other requirements, although people who had uploaded material more recently would be likelier to respond. The easiest way to identify individual Flickr members is to conduct a search and scan through the results, and a 'Tags only' search on 'meadows' presented several thousand results.

It was then a case of picking out any images with six or more tags, checking the member's personal page to make sure they had uploaded something in the last seven days, and adding their name to the list. The number of questionnaires sent out was settled at forty, intended to ensure a useful number of responses even if less than half replied.

The main ethical consideration was being open about what the questionnaire was for, so it was preceded in the emails by a short statement about its purpose. In the pilot the statement simply said that it was for research, but as a participant pointed out people were more likely to help a student as opposed to someone doing undefined research. It was thereby amended to specify a Master's degree. The compiled responses would eventually be included as an appendix with people's Flickr names left in. As regards anonymity none of the participants used their real or full names on the site so this was not deemed by the author to be a breach of privacy.

## 3.8 Low response issues

Once the research proper was underway a challenge arose in the form of a very low participant response; two weeks after the questionnaires were sent out, only four people had responded, three by replying to the email and another via the additional comments link. This was not a satisfactory amount for generating good qualitative data, so several techniques were used to try and increase response.

Firstly a reminder was sent out, again through Flickr mail, offering another chance to participate in the research. One possible reason for the sample population's lack of response could have been the lack of incentive, so an offer was included to make a prize draw for a large print of any single photograph on the author's page. It was also important to get responses in time for a full analysis so a deadline was given in the reminder, giving people two weekends (a popular time for uploading and browsing) in which to participate.

In an effort to increase the chances of better participation other Flickr members were sought out. A discussion page on a Pentax users' forum (Pentax User, 2008), discovered through a web search, included postings by several members. These people had shown themselves to be active in online discussions, and a look at their photo streams revealed that six of them also tagged in detail. These people were therefore sent internal emails inviting them to take part.

## 3.9 The search comparison

The third objective was to evaluate Flickr as a web-based image retrieval tool in comparison with a controlled vocabulary. Although the questionnaire touched on image retrieval, a comparison with a system using a controlled vocabulary would offer a more detailed picture of how well it worked for an external searcher. This meant turning to the commercial websites, ones that are designed to sell the images submitted to them. One of the most well-known and established of such sites is Getty Images. The company was founded in 1995 and the site holds both creative and editorial material that is used in newspapers, magazines, films, and other websites, under a range of licensing agreements.

Getty Images uses a controlled vocabulary called the Art and Architecture Thesaurus (AAT). The second edition of the AAT was published by Oxford University Press in 1994 on behalf of the Getty Art History Information programme[1]. Several sources were used including the Library of Congress Subject Headings, and although the project started in the eighties it was quickly apparent that the most common usage would be in computer-based systems. At the heart of the thesaurus are the descriptors, preferred terms that are usually nouns but can also be adjectives (modified descriptors) or verbal nouns. The descriptors are singular or plural depending on whether they describe objects (i.e. footballs) or an activity (football). The thesaurus is build around seven facets, from objects to associated concepts.

---

[1] The following information on the AAT can be found in Foskett (1996), pp. 412-416.

At this stage it would help to look more closely at how the two sites provide their metadata. While both have a title field for their images the approach is different; those on Getty are strictly descriptive (“Mixed breed dog looking away, close-up”), while Flickr users can be as cryptic and meditative as they like, or just leave the automatically-assigned file name (“IMG_4409”). For extra detail on Flickr there is an option to add a description, which some people use for fairly lengthy prose or even lines of poetry.

The AAT descriptors are included in Getty’s image metadata in the keywords field, and any words used in a search are mapped across to the closest matching descriptor; a search for ‘bravery’ will retrieve images with the keyword ‘courage’. The keywords field can also be used for specifics of content such as ‘one male only’, and a ‘Find similar images’ link brings up a list of keywords that can be selected individually or searched in combination.

Flickr allows a search by the full text of titles, tags, and descriptions, or a ‘tags only’ search. The tags themselves act as hyperlinks to all other images with the same tag, within one person’s collection (their ‘photostream’) or across the site. Members can join a host of special interest groups or add their images to a ‘pool’ based around a theme.

## 3.10 Designing the search comparison

The aim at this stage was to take the approach of someone searching for an image to fulfil a particular purpose. As mentioned already the images on Getty cater for a variety of clients who work in print, web-based and other mediums. In addition to the editorial photographs of particular events or people, the most common uses for images are illustrating an idea, getting a viewer’s attention, or providing decorative interest.

In order for the comparison to be fair the different boundaries of each site had to be taken into consideration. Getty holds images from professional or semi-professional photographers who are able to be present at key events or locations, and material submitted to them goes through a screening process for quality, while anyone can upload anything they want onto Flickr. Therefore judging the ability to find a photograph of a news item would not have been a valid criterion, and nor would the technical quality of the material.

Any retrieval system can be judged on some key areas: recall, relevance, and precision. Recall is the least applicable here, seeing that the two sites hold such a volume of material on any given subject that retrieving all the relevant material would not be desirable, although some variety is welcome. Morville is concerned with the way metadata captures the ‘aboutness’ of the item (2005, p. 125) and the ability to find images relevant to a subject is important. Relevance is hard to quantify in any meaningful sense, although precision (the ability to filter out irrelevant material) can be expressed in numerical form.

One problem with the comparison was that well-composed images with a clearly isolated subject are easier to describe and select, which puts Getty at a clear advantage given its semi-professional nature. Another problem was subjectivity; if the way an image is described is subject to an individual’s interpretation then so is its meaning to the viewer. As a consequence the author’s personal judgement of relevance in this stage of the research must be taken into account.

In addition to the criteria of relevance and precision, the effectiveness of a system can rest on its allowance for fluid browsing, the pursuit of a casual train of thought and a shifting of perception. If people and their ideas are the unreliable variable they can also hold the key to sharing information, and Flickr taps into what Morville calls the ‘power of gossip', leading to a kind of interactive foraging (p. 54). Bearing this in mind, the preferred approach to the comparison was a holistic one, using any available features and a variety of techniques to assess each system's effectiveness.

Having looked at the construction of the AAT it was possible to devise a suitable set of test queries. They would include entities, activities, abstract concepts, and combinations of these. Words with more than one possible meaning would also test the ability to handle homonyms.

In terms of quantitative data the two sites could be judged on how many pages had to be scanned through in order to find an acceptably relevant image, and how many irrelevant images appeared on the first page of results. In keeping with the holistic approach the overall experience of searching could be assessed for how easy and instinctive it was to use, including the browsing process within the searches.

### 3.11 Piloting the search comparison

Starting with specific entities, "Vulcan bomber" OR "Avro Vulcan" did not reveal any great disparity of results. Running this query as a 'Tags only' search on Flickr gave a choice of close-ups and distance shots to choose from on the first page (25 per page, out of a total 706 results), with just one irrelevant image. Getty gave one page of 20 quality images, all with the rare aircraft as the main object. Selecting the 'most interesting' filter on Flickr and looking closer at an image gave a link to a variety of groups and pools relating to historic and preserved aircraft and one specifically for the Vulcan. This group's pool provided a choice more on a par with Getty's, although it took three steps to get there.

Greater differences were revealed when searching for images of the Vulcan taking off. With 'Vulcan takeoff' as a tags-only search, Flickr gave 10 results, 6 of them relevant, while Getty did not respond at all to the query. The page of results from earlier did contain one image of a takeoff, and under 'refine this search', then 'keywords', then 'concepts' there was a link to 'arrival', which included the same takeoff image.

Turning to more abstract searches, 'resistance' met with some obscure results; Getty responded with all images bearing the keyword 'rebellion', so a few were relevant but many were not, although the 'concepts' link helped to refine results to areas like 'conflict' or 'adversity'. None of the immediate results on Flickr seemed to encapsulate the idea, but a brief browse through the pages and selecting the closest matches led to some groups who specialised in protest and demonstration images.

These pilot searches did start to reveal the strengths and weaknesses of the two systems, although a fairly standard query for a particular entity did not really show any great differences in precision or overall effectiveness. As a result the test queries were set up to locate images on the following subjects:

Tenacity
Red kite in flight (the bird of prey)
Red kite in flight (the man-made toy)
The loneliness of the long-distance runner

The aim in selecting these subjects was to test how well distributed classification did against a system with homonym control, and to judge the search and browsing effectiveness in response to subtle, complex ideas.

## 3.12 Reflective conclusion

The literature review worked well in giving an overview of the subject from a range of sources and it clarified the key terms in the working title. The reading also informed the search comparison aspect of the research by highlighting the importance of serendipity and browsing alongside accurate search retrieval.

Piloting the questionnaire was successful in making the questions more succinct and targeted at the objectives, eliminating overlap and reducing ambiguity for participants. However it did not help anticipate the problem of low response that emerged once the research was underway. Upon reflection the pilot could have been broadened to include Flickr members who were not already affiliated with the author, which may have identified this particular hazard at an earlier stage.

Because of the great wealth of literature available on search processes and behaviours the search comparison was well-informed at the piloting stage. Key changes that came out of the pilot comparison were the stronger emphasis on abstract concepts and a more challenging test of retrieving images of specific entities.

# 4. Results

This chapter comprises the results of the search comparison, in note form and as numerical data, and the compiled responses to the questionnaire.

## 4.1 Search comparison notes

*Red kite in flight* (the bird of prey): query “Red kite flying”

**Getty**: 9 images, 2 of a person flying a red kite, all the rest of the bird in flight (different poses & angles, one with rainbow in background, moon in another).

Browsing: selected ‘find similar images’ for one of the above, ticked subjects ‘animal’ / ‘gliding’ / ‘red kite’ / ‘no people’. Narrowed previous results to 6, all highly relevant with choice of styles (silhouettes against the sky, flapping & gliding poses, low tracking shots).

**Flickr**: 11 relevant on the first page, plus one of someone flying a red kite. Others on page featured this bird but too small or with other dominating elements (houses, trees sky etc., kite not main subject). 18 on second plus 4 of the toy, 18 on third with man-made kites featuring in the rest.

Browsing: Looked closer at 2 most relevant. Tag ‘milvus’ in second image gave 3 more by that person, all relevant, and 20 more in all public content, 14 relevant.
New search on ‘red milvus flight’ gave fully relevant first page with some variety.

* Searching ‘red milvus’ on Getty gave no results, just ‘milvus’ gave 2, slightly different.

*Flying a red kite* (the man-made toy): “Red kite flying”

**Getty**: Same 2 as before. “red kite person” filtered out all bird images but one (no people featured in this image yet ‘people’ appeared as a keyword).

Browsing: In most appropriate image, selected ‘similar images’, then subjects ‘kite’ / ‘flying’ / ‘leisure’. Gave 394 results, with first page of 60 images all of people flying kites, 2 of them red (plus the original image), 5 on second page. ‘Concepts’ to choose from within results included ‘fun’ / ‘enjoyment’ / ‘carefree’ / ‘togetherness’(all results showed more than person), ‘freedom’. Each of these contained at least one red kite on the first page, turned up 4 new ones between them.
Note: Searching these concepts individually meant going back to original results each time.

**Flickr**: Single image on first page (no more leads). 4 on second page (including close-up of a dragon kite), 3 on third, plus 3 featuring kites but not red ones. ‘Tags only’ increased number of images of man-made kites on first page, 2 fully relevant, 1 on second page, 4 on third. Some with people holding the kite, some with just the object (including a group of large novelty kites, one a red octopus).

Browsing: One of the images on the first page was a member of the ‘Kites & Kite Flying’ pool. Nothing relevant on first page, but internal search ‘red’ gave 9 relevant on first page, 10 on second, 6 on third. Emphasis on close-ups of the larger kites (unsurprising given the group) but some of people holding smaller, classic diamond-shaped ones too.

*The loneliness of the long-distance runner*(Looking for images that captured the sense of isolation in running vast distances alone): “lone runner”

**Getty**: Just 3 relevant results on first page (out of 60) but asked to clarify the search, so selected ‘one person’ and ‘running (physical activity)’; 4 on first page, 7 on second, 11 on third. Good variety of styles, angles and settings (track/urban/rural/beaches).

Browsing: In most appropriate image (lone runner on a dirt track, arid desert and mountains behind) selected ‘remote’ / ‘running’ from subjects, ‘one person’, and ‘loneliness’ / ‘solitude’ from concepts. Gave 46 results, 30 relevant, with wide range of styles.

**Flickr**: 10 on first page, 11 on second, 11 on third. Some variety, dominated by beaches. ‘Tags only’ reduced results to just 7 with 4 highly relevant, some variety (2 beaches, one urban, one beach sunset).

Browsing: First image (highly relevant) linked to pools ‘Emptiness’ (no more relevant images turned up by searching ‘runner OR running’ within the pool) and ‘Alone and lonely’ (same internal search as above turned up one relevant image, itself part of ‘running’ and ‘solitude’ pools). ‘Solitude’ pool (same internal search as above) gave 4 relevant on first page, 1 on second 1 on third). Search ‘lone OR solitude’ within ‘Running’ pool gave 7 results, one highly relevant (long mountain road, lone runner).

*Tenacity*: “tenacity”

**Getty**: First page of 60 results had 42 relevant images, variety quite low with several near-duplicates. Results dominated by subtle variations on shots of athletes straining or looking focused. 2nd and third pages likewise.

Browsing: Picked one from third page (boy doing pull-ups), ‘find more like this’ selected ‘expressing positivity’ / ‘people’ from subjects, ‘effort’ / ‘confidence’ / ‘determination’ from concepts. 34 images on first page expressed tenacity in some form.
Note: The keyword ‘tenacity’ never appeared as keyword in any of images, only concepts like those mentioned above, or ‘persistence’.

**Flickr**: 12 relevant on first page, 13 on second, 11 on third (dominated by plants growing in unlikely places, mostly titled with tenacity, some tagged as well). ‘Tags only’ gave 9 on first page, 8 on second, 3 on third. More variety than full-text search with more of people and animals (looking determined / clinging on).

Browsing: Selected individual photos looking for other tags or group pools that might contain more on the subject, looking out for highly commented-on items. From first page of ‘tags only’ search, one of a weed breaking through tarmac was part of group pool ‘Voices in the wilderness: a prayer for wild things’. 9 relevant out of first page of 30, internal search on ‘tenacity’ gave 4 images, 3 relevant (included original image that led into the group).

Selecting several images from first page of ‘tags only’ search, several also tagged with ‘fortitude’ and ‘determination’. Tags only search of former proved disappointing, with one of the images from earlier the only relevant one on the first page (very senior person walking with a stick). ‘Determination’ better, with 9 images displaying tenacity on first page, 11 on second, 8 on third. Good variety of different people shots, animals, and some plant-life.

## 4.2 Relevance ratings

| Search term | Getty | | Flickr | | |
|---|---|---|---|---|---|
| *Red kite in flight (bird):* | 78% | | 46% | | |
| Page 2 | | | 75% | | |
| Page 3 | | | 75% | | |
| | | | | | |
| Browsing: | 100% | | 100% | | |
| Page 2 | | | 70% | | |
| Page 3 | | | 100% | | |
| | | | | | |
| | | | | | |
| *Flying a red kite (toy):* | 22% | | 4% | Tags only: | 8% |
| Page 2 | 67% | | 17% | | 4% |
| Page 3 | | | 13% | | 17% |
| | | | | | |
| Browsing: | 3% | | 38% | | |
| Page 2 | 8% | | 42% | | |
| Page 3 | | | 25% | | |
| | | | | | |
| *Loneliness of the long-distance runner:* | 5% | | 42% | | |
| Page 2 | 7% | | 46% | | |
| Page 3 | 12% | | 46% | | |
| Page 4 | 18% | | | Tags only: | 57% |
| | | | | | |
| | | | | | |
| Browsing: | 65% | | 20% | | |
| Page 2 | | | 17% | | |
| Page 3 | | | 4% | | |
| Page 4 | | | 4% | | |
| Page 5 | | | 14% | | |
| | | | | | |
| *Tenacity:* | 70% | | 50% | Tags only: | 38% |
| Page 2 | | | 54% | | 33% |
| Page 3 | | | 46% | | 13% |
| | | | | | |
| Browsing: | 57% | | 75% | | |
| Page 2 | | | 4% | | |
| Page 3 | | | 38% | | |
| Page 4 | | | 46% | | |
| Page 5 | | | 33% | | |

## 4.3 Questionnaire responses

How do you select your tags?
(what elements of a photo do you use to select them?)
**Manganite:**Place, date, camera data, topic, colors, postprocessing
**Onnufry:** I ussually tag information about (each if applies):
- location name (country, city and/or region... "paris", "sahara desert")
- location type (e.g. "city", "village", "highway", "oasis")
- object(s) name (given name, e.g. "eiffel tower")
- object(s) type (e.g. "tower", "river", "building", "man", "couple")
- object(s) dominating characteristic (e.g. "steel", "red", "smooth", "happy", "fast")
- action description (e.g. "sleeping", "hitchhiking", "cooking")
- genre and theme (e.g. "architecture", "landscape", "portrait", "tourism")
- lense used (since it's not stored in EXIF)

**duncan:** I tag the most obvious attributes of the photo - what does it depict? where is it? what keywords might someone else search for to find this photo? etc...
**Effervescing Eleaphant:** I generally tag places, and maybe what's happening in the picture.
**Badly Drawn Dad:** I try to tag as fully as possible. I use physical descriptions of the subject and its surroundings. Sometimes I add abstract or humorous tags for comic effect.

Why do you tag your photographs?
**Manganite:**For find certain pictures on my own and to have my pictures found by others. I also use them via API [Application Programme Interface] to select certain pictures for my own homepage
**Onnfury:** Tags help people finding photos. Tags help photos finding people.
**duncan:** 1. So I can find them myself later
2. So other people might stumble across them when searching via Flickr
3. to improve their SEO rating on Google etc so people searching outwith Flickr might also find them
**Effervescing Eleaphant:** I allow creative commons on most photos (all those not specifically of people) and get a buzz seeing them appear on Wikipedia and other places. I don't believe anyone would pay for my photos so it's fun.
**Badly Drawn Dad:** One of the reasons for posting photos on Flickr is to show off. I want the maximum number of people to see my pictures and possibly add a comment. I am pleased when my photos get favourited or attract many views.
Another reason is that Flickr is a resource: if I want to see a photo of Tower Bridge at night I know that by searching with the right tags I'll get ten thousand appropriate pictures!
(BUT I belong to the Guess Where London Group, where the point is NOT to tag photos until they're guessed.)

Would you describe adding tags as a positive experience?
**Manganite:** Yes, definitely
**Onnfury:** It's just something that needs to be done. Neither positive nor negative experience here, apart from situation when you've got no idea how to tag something...
**duncan:** Yes, mostly
**Effervescing Eleaphant:** Yes
**Badly Drawn Dad:** It's a bit of a chore, but untagged photos are almost useless except to their owner.

### What would make tagging easier or more enjoyable?

**Mananite:** automatic addition of synonyms and multilingual tags
**Onnfury:** By technical means it depends on the application and it's user interface and it's a very long story, really. And that's only about "easier" part, nothing to say about "more enjoyable".
**duncan:** Some sort of system that could auto-suggest similar tags that may be relevant. Not having to tag different variations of the same word, e.g. "cake", "cakes" etc.
**Effervescing Eleaphant:** It's rather labourous, but it's fine. Might be useful to have a drag and drop facility from tages used before.
**Badly Drawn Dad:** Geotagging: when I add a photo to the map it would be useful if the name of the country, town and street could be added automatically. Predictive text based on my existing tags might be helpful.

### Do you use Flickr to search and browse other people's photos?

**Manganite:** Yes, I do.
**Onnfury:** Yes. Searching photos regarding travel (places I've been to or will be), home (places, events), work (cc photos if I need something).
**duncan:** yes
**Effervescing Eleaphant:** Yes
**Badly Drawn Dad:** Yes, that's what they're for

### If so, how accurate are the results of your search?

**Manganite:** Most time sufficient, but I guess a lot of good pictures are not tagged so never be found...
**Onnfury:** Often cluttered. Many people tag photos inaccurately. Even more people don't bother to tag at all. But I ussually get what I need.
Some people tag photos in a non-english language only, that's not helping here too.
**duncan:** Reasonably
**Effervescing Eleaphant:** Pretty good, but some irrelevent photos which I skip past.
**Badly Drawn Dad:** [If so, do the results reflect their tags accurately?] Usually, but sometimes (hilariously) not.

### How could the quality of Flickr tags be improved?

**Manganite:** again, automatic addition of synonyms and multilingual tags
**Onnufry:**
Nothing to change really. Simple tags are good enough, even with their flaws. And it is likely impossible to change that.

Like... how to force people to tag? Or how to force them to tag in one language?...

One feature that would be nice: built-in, transparent translator. Let's say, I search for "book". Searching system should automatically search for both "book" and it's dictionary translations to other languages - "buch", "libro", "livre", "kniha", "العربية" or "□□" :D
**duncan:** Don't know
**Effervescing Eleaphant:** Pretty happy :o)
**Badly Drawn Dad:** Some sort of 'fuzzy search' so that different spellings or hyphenations would be equivalent.

# 5. Analysis of the results

This chapter comprises an analysis of the search comparison between Getty Images with its controlled vocabulary and Flickr's tag-based system, as well as the questionnaire responses from Flickr members. Some of the assertions made in the literature about the effectiveness of distributed classification (which tags are part of) will be discussed in the light of results. The low response issues around the questionnaire will be addressed, although the answers that did come back were of sufficient detail to help appraise the different things written about what such a system is like for its inhabitants, and indicate how effective they think it is.

## 5.1 The search comparison

Comparing the two systems offered an opportunity to test out some of the statements made on the blog postings and in the journal articles. For example, would tags prove to be such a disaster for targeted searching as Merholz thinks (2004), or would Mejias and Shirky be proved correct in their assertion that distributed classification 'works rather well' (Mejias, 2004) despite its lack of structure and negotiated meaning? Indexing writers such as Foskett and Enser both advocate the value of browsing, and Merholz concedes that a tag-based system is ideal for it, but a detailed account of browsing through the two systems offers the chance to see whether Flickr has any demonstrable advantage.

Each of the test queries was used to judge the overall precision of search results, the facilities for browsing, the degree of specificity allowed by the system, and the variety amongst results. A judgement of relevance is dependent on the end purpose of any given search, so it is possible for an identical query and set of results to have varying relevance, as in the case of the different types of red kites (birds and toys). Relevant images were totalled per page of results (60 images per page in Getty, 24 in Flickr), and up to four pages were looked through depending on how satisfactory the results were. This was intended to represent how they would be used in the real world, and the total recall number was not looked at, as scanning through over a thousand images would not be practical.

A person looking for an image does not always know precisely what they want (Enser, 200, p. 206), so good browsing facilities let them pursue an aspect of the subject that might arise form an initial search (Foskett, 1996, pp. 25-26). Good metadata can also help to formulate a new search or refine a set of results; this also relates to specificity, the degree to which a query can be narrowed and refined. A certain level of variety can help someone make a choice without looking through too much material, and photographs are as much about lighting, angles of view, and positioning, as they are about their physical subject.

## 5.2 Quantitative analysis

In order to collate some quantitative data on the precision of the respective systems, the search results were broken down to the number of relevant images. By placing the number of images picked out from any given page over the total number of images on the page, it was possible to represent precision as a percentage. It was possible to draw up an average percentage where pages were looked at in sequence.

Looking over the collated precision ratings it was possible to see several patterns emerging. Getty scored higher in the first two subjects on red kites ( the birds, then the toy), and again in the last subject (tenacity). The difference was greatest in the second subject, flying a red kite, but only after the search was amended. In a break of pattern however the lone runners search put Flickr significantly higher, even after the search on Getty was clarified with extra limitations.

One of the clearest and more striking patterns was centred on the use of the 'tags only' feature. Where a Flickr search was limited to just tags it increased the relevance of results in only one instance (lone runner), but in all other cases it actually decreased the precision. Meanwhile the precision within the browsing results showed a reverse of the search stage, apart from the first example where it was close to an even match. These patterns will now be examined in closer detail.

## 5.3 Precision and specificity

The query 'red kite flying' was used as the initial search statement for both the bird and the man-made toy. Out of Getty's 9 results, 7 were of the bird and the other two of the toy, and amending the search to 'red kite person' filtered out all of the birds but one. A scan through the first 3 pages of Flickr's results averaged out at 65% relevance for the bird and 11% for the toy (as opposed to Getty's 78% and 67% respectively), which discounted any instance where the object was either too small within the frame or dominated by another element (e.g. buildings or trees). In this instance the total of the images recalled is an issue, because while Getty's relevance ratings look impressive it was within a much smaller data set, where accuracy is much easier to achieve, and even then a clearly irrelevant result still managed to get through. No people featured in the bird picture in question, yet 'people' still featured as a keyword. This shows how inaccurate metadata can still find its way into a controlled and mediated system.

Entering 'lone runner' in Getty gave just 3 relevant results on the first page of 60 images (5%), although there was an option to clarify the search; selecting 'one person' and 'running' improved the average relevance to 12% over 3 pages. The ability to specify in this way was impressive, but there were still very few images that captured the subject. Flickr did much better here with 45% of the images from 3 pages being relevant, while a 'tags only' search reduced results to just 7 images, 4 of them highly relevant. When it came to searching 'tenacity' the first page of Getty's results was enough, with a relevance of 70% and no obvious inaccuracies. Flickr managed an average of 50% across 3 pages, which was acceptable.

It was here that an unexpected phenomenon presented itself: searching with just tags actually decreased the relevance of results, and a significant number of the images did not represent the idea of tenacity at all. Likewise in the second red kites example the 'Tags only' search reduced an already low relevance rating. This decrease in precision could indicate that more attention and thought is given to titles (the majority of 'tenacity' images had been given exactly that name) than the other metadata fields. In the last example there were whole sequences of images of a cooking event from one person, all tagged with tenacity. Cases such as this serve as a reminder that tagging tools can offer just as much opportunity for poorly considered, hastily applied metadata, as they do for good tags.

## 5.4 Variety

Remaining with the example of the 'tags only' search on tenacity for the moment, it is worth noting that it did increase the variety amongst the relevant images. The full-text search was dominated by plants or trees growing in unlikely places, whereas limiting it to tags introduced more of people and animals, looking determined or clinging on. In contrast the variety in Getty was quite low with several near-duplicates, dominated by shots of straining athletes.

Amongst the ‘lone runner’ results Getty displayed a much better variety of styles, angles, and settings, Flickr less so with mainly beach shots, although one urban setting as well. In the two different red kite searches both sites gave an acceptable level of variation of styles and angles. There was slightly more in Getty’s small set of results which had silhouettes, flapping and gliding poses, and a low tracking shot. With the man-made object Getty only gave the two images, one of just the object and the other with someone holding it. Flickr was the same but with a few large novelty kites in there as well, one advantage of having a larger data set.

## 5.5 Browsing

Once the initial search had been brought to a conclusion the browsing process began. The standard method was to find the image that encapsulated the subject best, and use any details or features in the metadata to follow the idea. In the case of the first red kites example, one of the more striking bird shots in Getty contained a ‘find similar images’ link. This presented a list of subjects and concepts, and selecting ‘animal’, ‘gliding’, ‘red kite’, and ‘no people’ narrowed the previous set of results to just 6 images, all fully relevant. Because of the very low number of initial results this was more filtering than browsing, although this technique had more effect in the second example, where the subjects like ‘kite’, ‘flying’, and ‘leisure’ and the concepts ‘fun’, ‘free’, ‘enjoyment’, and ‘togetherness’ gave pages full of people flying kites together, turning up a total of 4 new images of red ones across two pages. It is not clear why these were not present in the original set of results (or why selecting ‘red kite’ as a subject had a similar limiting effect), but the detail and clarity of the metadata certainly allowed a new idea to be comprehensively explored.

For the red kite examples in Flickr two slightly different browsing methods were employed. In one of the most relevant bird images the tag ‘milvus’ gave 3 more by that person, and 20 more in the public content (14 relevant), and a new search on ‘red milvus flight’ gave a fully relevant first page. In the second case one of the images was part of the ‘Kites & Kite Flying’ pool, where the internal search ‘red’ gave a lot of large kites in close-up plus a few more classic types, with an average relevance of 35% (as opposed to Getty’s 4%). In the ‘lone runner’ and ‘tenacity’ examples it was the group pools that gave the most leads; using the internal search ‘runner OR running’ in the ‘Alone and lonely’ pool gave one new image on the subject, which itself led to the ‘solitude’ and ‘running’ pools. For tenacity, a weed breaking through tarmac belonged to a pool called ‘Voices in the wilderness: a prayer for wild things’ where an internal search on ‘tenacity’ gave 4 images, 3 relevant, including the original image that led into the group.

While the ‘tags only’ search on tenacity was disappointing in terms of immediate results, several of them were also tagged with ‘fortitude’ (no new results) and ‘determination’, which gave 3 pages that were 39% relevant to the subject, with a good variety of plant, animal, and people shots. Over in Getty a shot of someone doing pull-ups led to the subjects ‘expressing positivity’ and ‘people’, and concepts such as ‘effort’ and ‘determination’; 34 images on the first page expressed tenacity in some form.

Tenacity never appeared as a keyword in Getty, only concepts like those mentioned above or ‘persistence’. In this case it did not present a major problem, but as the resistance/rebellion example from the pilot illustrated, the choice of preferred terms in a controlled vocabulary does not always match with the way its users think. People on Flickr can tag their photos with the word tenacity if they wish to, which helps to support the assertion by those such as Vander Wal (2005) and Matusiak (2006) that distributed classification allows a closer fit with someone's own understanding.

Both systems gave the opportunity to follow an idea and discover new ones in their own way. Getty's use of associated subjects and concepts in a tick-list is pleasingly methodical to use, rigidly logical, and if the searcher knows what they are looking for they can specify to a very close degree. At times though it feels too rigid, particularly when investigating a subtler mood or concept. It is here that Flickr's interlinked networks of special interest groups is at its strongest, tapping into the interpersonal connections identified by Morville (2005). Tags on their own often produce some baffling and unsatisfactory initial results, but in combination with the group pools they offer a random yet subtly sophisticated tool for discovery.

## 5.6 The questionnaire

In the methodology chapter the poor response rate was mentioned, and some of the strategies employed to increase participation. However when the cut-off date for collecting responses came round there were still only four, so the decision was taken to include one of the responses from the pilot. This person was an active Flickr member and tagger, like the others from the research proper, and the questionnaire they had been sent was the closest to the final version (to the extent that all the answers were there). The reasons for the low response will be speculated upon later, but before that the questionnaire data will be analysed in full.

People selected their tags using a mixture of spatial and temporal data, featured objects or activities, abstract concepts such as mood and genre, and technical details relating to the camera or post-production techniques. Everyone mentioned location in their answer, one person includes the date, and one person includes the type of location and objects as well as the names. Nearly everyone aimed to tag the most salient feature of their images, whether they expressed this as the topic or obvious attributes. The term abstract was used once, and all respondents used some kind of attributed term in their description. One person also talked about trying to predict the terms people would use to find their image. So far this seems very close to the various levels of metadata described by Eakins, and in some ways it is even more detailed due to the inclusion of camera and post-production details.

Asked why they tag their photos, getting them found by other members was cited as a reason by all five. Two people said it was also to help themselves find them later (important if they have hundreds of images stored on their photostream). One of these people uses tags in conjunction with an Application Programme Interface (API) to select them for a separate homepage, while the other participant uses them to improve the photo's findability in web search engines like Google.

Where the desire to help others people find the photos was elaborated upon, the reason given was some form of prestige; either the immediate 'buzz' of having work commented on and marked as a 'favourite', or seeing a photo used on Wikipedia. One person saw tags both as part of making Flickr a functioning resource, and as a means to playing a 'guess where' game in a group (members have to guess where a photo was taken and be the first to add it as a tag). In a startling echo of Ranganathan's laws that state 'for every reader, a book… for every book, its reader' (Koehler, 2004, p.401) one answer was simply 'Tags help people finding photos. Tags help photos finding people.' Even considering that the participants represent the naturally cooperative minority, and their own aims of gaining prestige, their awareness of tagging's collective value stands against Merholz's idea that all benefit to the wider community is entirely residual.

Three out of the five participants answered 'yes' when asked if tagging was a positive experience; one added 'mostly', another 'definitely'. The other two saw it neither as positive or negative, just necessary, although these people did describe it as a chore and a challenge respectively. Bearing in mind the previous answers it is possible to say that people are happy to add tags if they can see potential value to themselves and others.

Suggestions for making tagging easier or more enjoyable included 'automatic addition of synonyms and multilingual tags', automatic suggestion of alternative words and variations (e.g. plurals), and 'geotagging' that would work in conjunction with placing the image on a map. There were also technical, time-saving solutions like predictive text or drag-and-drop based on someone's existing tags. No-one thought that it could or should be made more enjoyable, barring the earlier comment about the 'guess where' games. However all the suggestions above would increase consistency in addition to convenience for taggers, while still maintaining the essentially free nature of the classification system.

All participants use Flickr to search and browse other people's photos. The only person to elaborate said that they searched for places they might visit (or had visited), places and events of interest, and material under the Creative Commons license for professional use. As regards the accuracy of results there was an overall consensus that it was sufficient, often cluttered with irrelevant material but still able to find things. Someone made the comment that many good photographs will never be found due to being poorly tagged, while another person lamented the material only tagged in a non-English language. These answers suggest that while Flickr's members find some aspects of the classification system frustrating, it still delivers enough of what they want.

When it came to suggesting ways to improve the quality of tags on Flickr one person reiterated their call for automatic addition of synonyms and other languages, and a 'fuzzy search' to cover variations in spelling and hyphenation was also proposed. One participant accepted that people could not really be forced to tag or use the same language, but suggested that a built-in translator would help. Three people either didn't know any ways to improve the system or were just happy with it, including the one who suggested the translator. Much like Guy and Tonkin (2006) the five participants had a few ideas for enhancing tagging, particularly in the area of consistency, but they still accept it in its existing state as a functioning tool. As one of them put it, 'Simple tags are good enough, even with their flaws. And it is likely impossible to change that'.

## 5.7 Why did so few participate?

The reasons for the low response rate can only be speculated upon, but one possible reason is the speed at which Flickr is evolving, along with the rest of the social web. The four years in which Flickr has been running is an eternity in the digital environment, and while tags remain the core description tool for individual images on Flickr its members' focus may well have moved on, along with any desire to debate the subject. Even in a questionnaire focused on tags, membership of a group was mentioned as part of an answer, and in the search comparison the group pools proved key to successful browsing. It is also likely that Flickr members travel a lot and many people are away from their keyboards during the summer period, but even long vacations couldn't entirely account for the lack of response. The more plausible scenario of the focus shifting to personal interaction will be discussed further in the conclusion chapter.

# 6. Conclusion

The aim of this research was to evaluate Flickr's tagging-based system as a tool for describing, organising and retrieving web-based images. A measure of its success is the depth to which the evaluation was achieved, and the data generated did help to test out the claims and assertions made about distributed classification. The main weakness lies in the low response rate to the questionnaire and the lack of any exchange between participants in the discussion space that was set up, even after extra people were contacted and an incentive given. This was a disappointment, and having more participants could have produced a greater ranger of opinions, but the results that did come back were of high quality and interest.

## 6.1 The research process

A core strength of this research was the level of insight into the subject gained from the literature review and both parts of the method, particularly the search comparison. Newly published articles and blog entries were caught and commented upon, even ones that appeared after the initial plan had been submitted, which helped to keep the focus current and relevant to what is happening with online image collections in a wider context.

The issues that were picked from the literature were fed into the design of the questionnaire, such as whether distributed classification helps involve users (Fox) and makes sense to them (Chowdhury & Chowdhury). The questionnaire was also amended effectively on the basis of the pilot responses, removing areas of overlap and ambiguity. As a result the answers to the final version showed a much better level of comprehension.

Quality of the data arising from the search comparison indicated that it was highly successful in its purpose. It was designed to test the effectiveness of both systems in targeted searching and casual browsing, and the picture built up was both complete and revealing. It may have been possible to conduct and present the comparison in a more methodical way, but that would have risked losing the instinctive, holistic approach that was intended to replicate real world use.

## 6.2 Reflection on the results

Answers to the questionnaire showed a generally high awareness of what makes an image easier to find. Key details like location and date were routinely used, as well as conceptual elements such as overall theme and style, although not every person said they used all of these details. One area where Flickr members have used their initiative and gone beyond the traditional areas of image indexing is the inclusion of technical details on the camera, lenses, and post-production techniques used; because these details are included as tags it makes it possible to search by them, although a lack of prescribed format could make it unpredictable.

It should be acknowledged that the five people who participated when 42 chose not to are likely to represent the more benign and willing portion of Flickr members. Despite this there was a surprising awareness that tags could have a communal benefit as well as helping them organise their own photos. The willingness to increase an image's visibility to other members and web users was partly influenced by a desire for recognition, but not entirely. Tagging was regarded by some as necessary chore but they were happy to do it; alongside a obvious striving for accuracy, such answers indicated that the benefit of tags can be truly universal.

In terms of the quality of tags and their usefulness in searching the overall answer was that they were good enough for their purpose, if occasionally frustrating. This fits with most of the professional literature and blog entries. What did not fit so closely were the calls for prompts and suggestions to help improve consistency when adding tags, as well as the automatic addition of synonyms and translation. Guy and Tonkin have voiced reservations about trying to control tags too heavily, but if taggers just want a few prompts in areas like spelling then it would be possible to improve quality, without compromising the essential fluidity.

What the search comparison demonstrated is that controlled vocabularies and distributed classification are attempts to address the image indexing challenge in different ways, each with their own advantages. The strength of the Getty site lies in its capacity to specify with accuracy; having control over descriptive terms and a full structure of subjects means that ambiguous terms are simple to deal with. Where the search subject was less well defined and more subtle, the rigidity of controlled terms can feel constraining, particularly when the searcher's preferred term does not conform to that of the vocabulary.

In the case of Flickr the comparison showed that tags are not necessarily the most important part of the description; an accurate title can often do just as good a job. As the participants in the questionnaire pointed out, the site's performance as a search tool is acceptable if not perfect, and its ease of navigation helps to gloss over much of the inaccuracy. The instinctive nature of browsing within a set of search results is what closes the gap with a more traditional system. This is achieved by the potent combination of tags and the groups, allegiances of interests or themes that tap into shared ideas. The function is quite similar to the structure of a controlled vocabulary.

## 6.3 Further research

It was the groups on Flickr that caught the author's attention at numerous points in the research process: they were mentioned in the questionnaire, proved an important part of the browsing section in the comparison, and they provide an indication as to why so few people showed an interest in talking about tagging. At the time of writing Flickr has been running for four years, and it appears to have evolved significantly in the way its collection is organised. Tags are still important but only in the sense that they enable members to form groups. It was not uncommon to find photos with only two or three tags but links to six or more group pools on related subjects. This would be the natural focus of any further research into distributed classification and images.

There is also a growing trend for external cultural organisations to tap into Flickr. The Library of Congress has made a collection of historic photographs from American life available on the site in a project called 'The Commons' (Flickr, 2008). The images can be tagged and commented on by Flickr members, while the Philadelphia Museum of Art has created a part of its website where digital images can be tagged by any web user (Philadelphia Museum of Art, 2008). Projects such as this could represent a new online use for distributed classification beyond casual sharing of photographs, a possible area of exploration for the information profession.

# Appendix 1: Dissertation plan

**Working title of the dissertation**

An evaluation of Flickr's folksonomy from the perspective of its members, and as an image retrieval tool in comparison with a controlled vocabulary.

**Brief review of the literature and list of references (200 words maximum)**

The emergence of digital camera technology and the spread of broadband have made a previously unimaginable amount of images available online. Creating image metadata has always been an ambiguous, labour-intensive process (Eakins, 1998), and digital advances have tended to run ahead of approaches to organising information.

Tagging, the assignment of simple keywords by whoever uploads an image, appears to offer a solution. This process of user-indexing has been labelled as a folksonomy, mob-indexing, tagsonomy, distributed classification, social classification, or ethnoclassification. Folksonomy is perhaps the most widely-used and evocative, although its allusion to taxonomies is misleading as there is no strict hierarchy (Matusiak, 2006). The key is that the process is collaborative, the descriptive terms being chosen by the users of a system and shared amongst them.

While most commentators are quick to point out that tagging is often sloppy and unsophisticated when compared to the semantic links of a traditional scheme (Chowdhury, 2007, p. 108), there is recognition that folksonomies empower web users and make sense to their community. All communities, however, are diverse in makeup (Lilley, 2006), and the experience of the individuals who participate in tagging-based image collections has yet to be properly addressed by the information profession.
200 words

**Research Design (300 words maximum)**
The Flickr website started in 2004 and is the most prominent example of an online image collection based on user-assigned tags and free-text descriptions. It is designed for personal organisation of digital photographs, and for sharing them with other Flickr members. Getty Images is an online collection of creative and editorial stock, designed for sale and using their own thesaurus as a controlled vocabulary for keywords.

**Overall aim**
Evaluate the success of Flickr's folksonomy-based system in describing, organising, and retrieving web-based digital images, in comparison with Getty Images.

**Objectives**

1) Identify the most positive and negative aspects of tagging as an image description and organisation tool, from a tagger's perspective.
2) Evaluate the success of tags in retrieving online images, from the perspective of Flickr members.
3) Analyse the effectiveness of Flickr's folksonomy as a web-based image retrieval tool, in comparison with a controlled vocabulary.

**Method**

- Open-ended questionnaire using the Flickr mailing system, based on a sample of users selected from a 'tags only' search (Qualitative, objective 1).
- Open-ended questionnaire to a separate set of Flickr members, selected from a 'full-text' search (Qualitative, objective 2).
- Initiation of a discussion between all questionnaire respondents, using the comments feature under a designated photo on the author's page (Qualitative, objectives 1 & 2).
- Critical comparison of search effectiveness between Flickr and Getty Images, using a common set of themes (Mixed-method, objective 3).

**Sampling strategy**
Recipients of the questionnaires will be selected from the top pages of search results onwards without any regard to their images' characteristics, to reduce the risk of bias. The two searches are intended to distinguish between those members who tag, and those who use free-text descriptions as their main form of metadata. The search comparison will feature both abstract and specific themes.

297 words

www.flickr.com
www.gettyimages.com

**Reflection on Ethical Research issues  (100 words)**
If any images are displayed as part of the dissertation it will be necessary to gain permission from their owners. They will be given the option to remain anonymous or be credited by name, having been made aware of the context in which their images appear. Questionnaires and discussions will need to be accompanied by an assurance of anonymity in published results and a clear statement of the research objectives. The discussion should also be regularly monitored for abusive comments. It would be considered courteous to notify the Flickr administrators of the research methods and objectives.

96 words

**Timetable:**

| | |
|---|---|
| October – November: | Formation of the research topic based on several journal articles. Class-based learning of research design and methods. |
| December: | Further reading, forming the start of the literature review. |
| January: | Finalisation of research topic. Dissertation plan. |
| February – April: | Draft of the Literature Review. Questionnaire design. |
| April – May: | Piloting of the questionnaires & conducting the comparative search analysis. Sending questionnaires and initiation of the discussion. Writing the Methodology chapter. |
| June – July: | Compiling and analysing findings. Writing the Results Analysis chapter. |
| July - August: | Final editing and write-up, including the Introduction and Conclusion chapters. |

## Appendix 2: Examples from Flickr

The top half of a page of search results

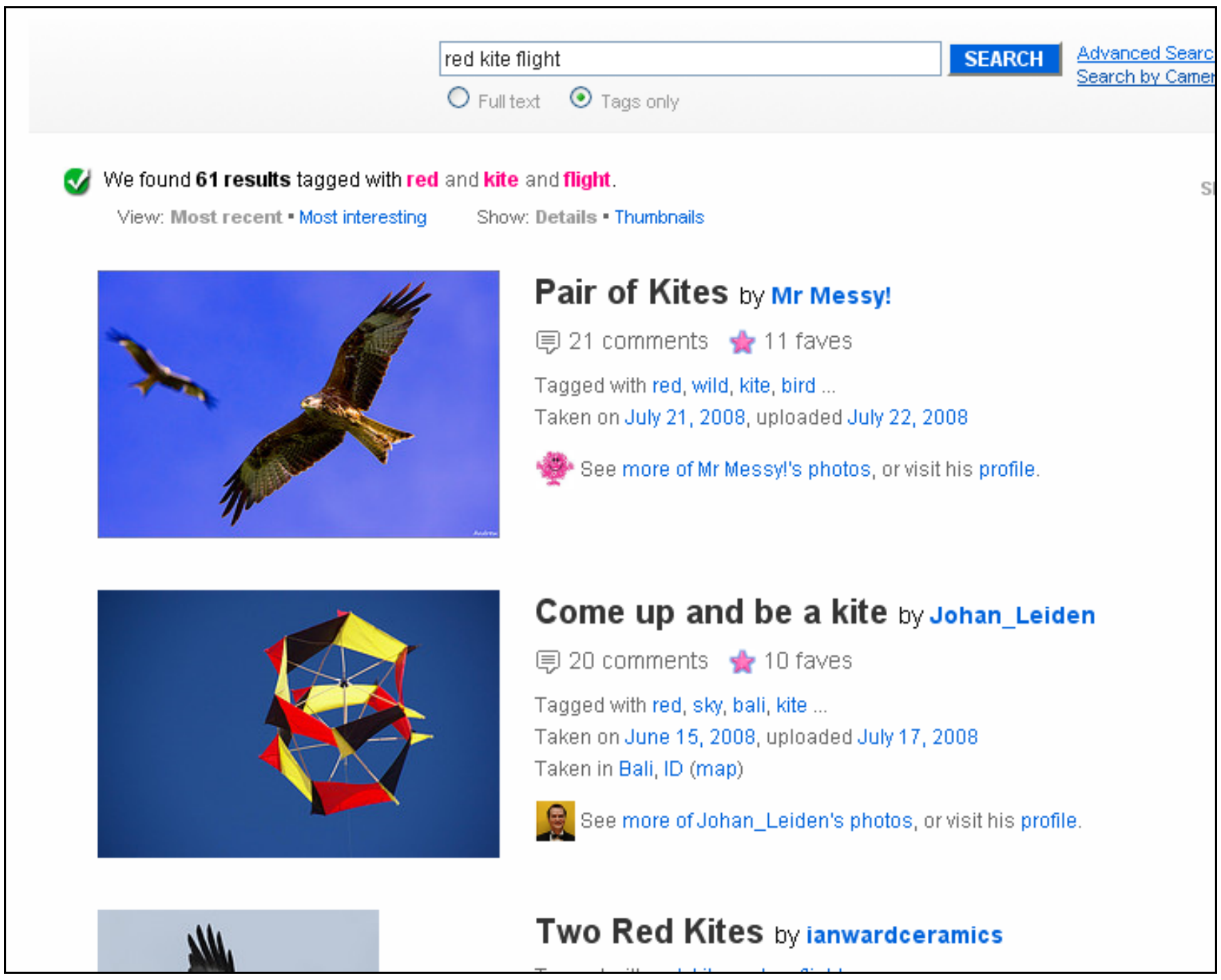

A tagged image from the author’s photostream

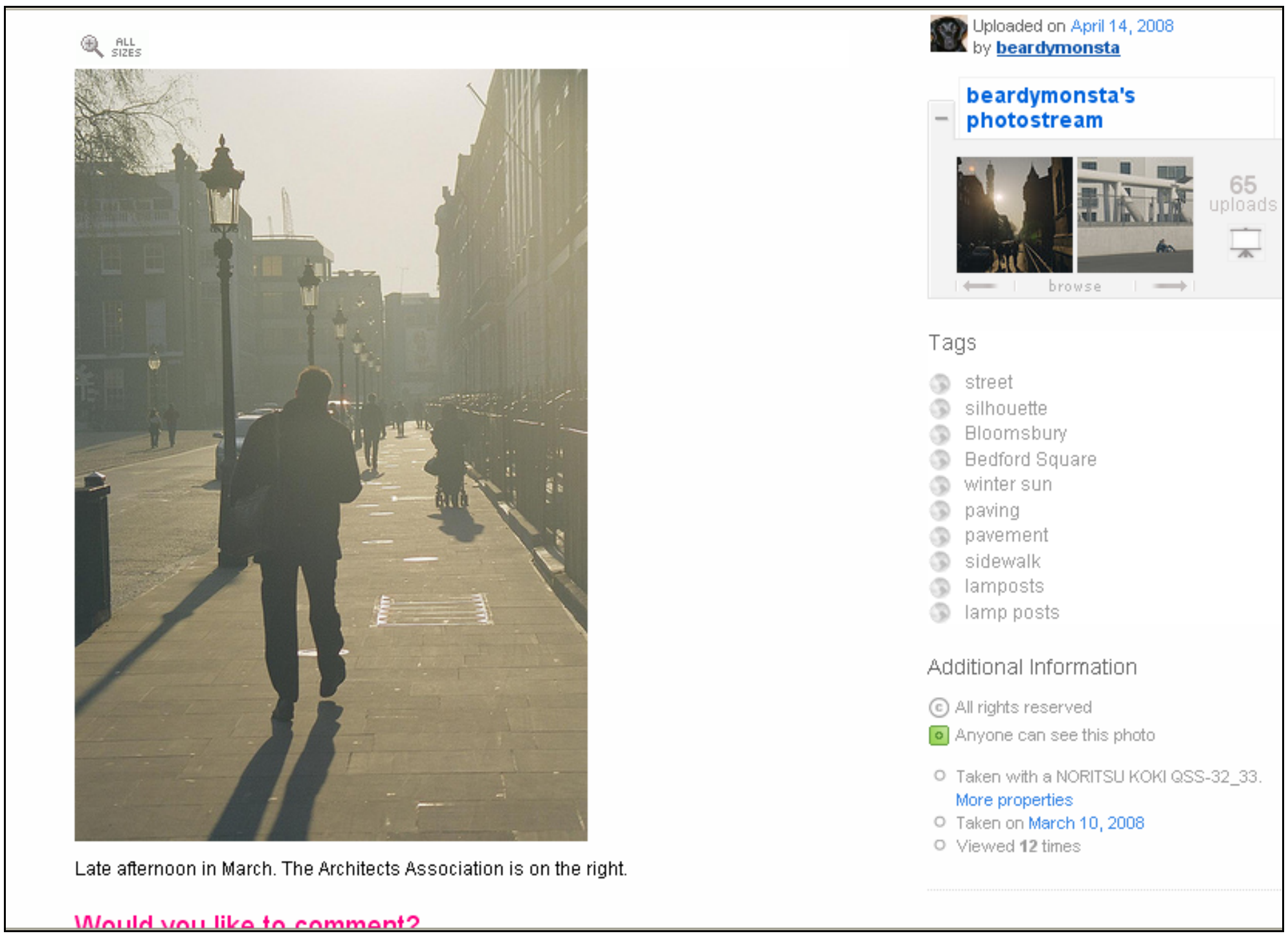

# Appendix 3: Examples from Getty Images

A set of search results

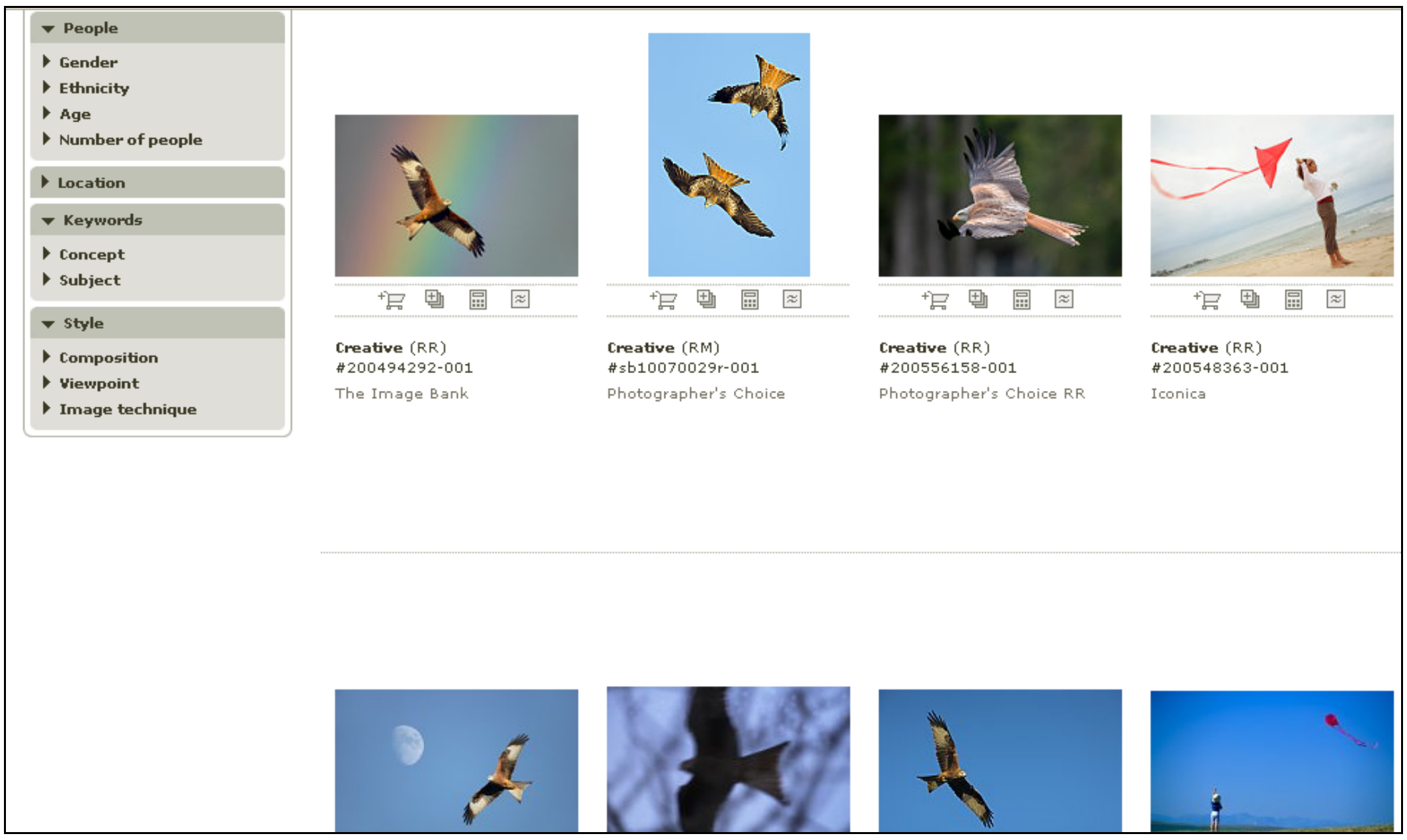

An image with accompanying metadata, including keywords

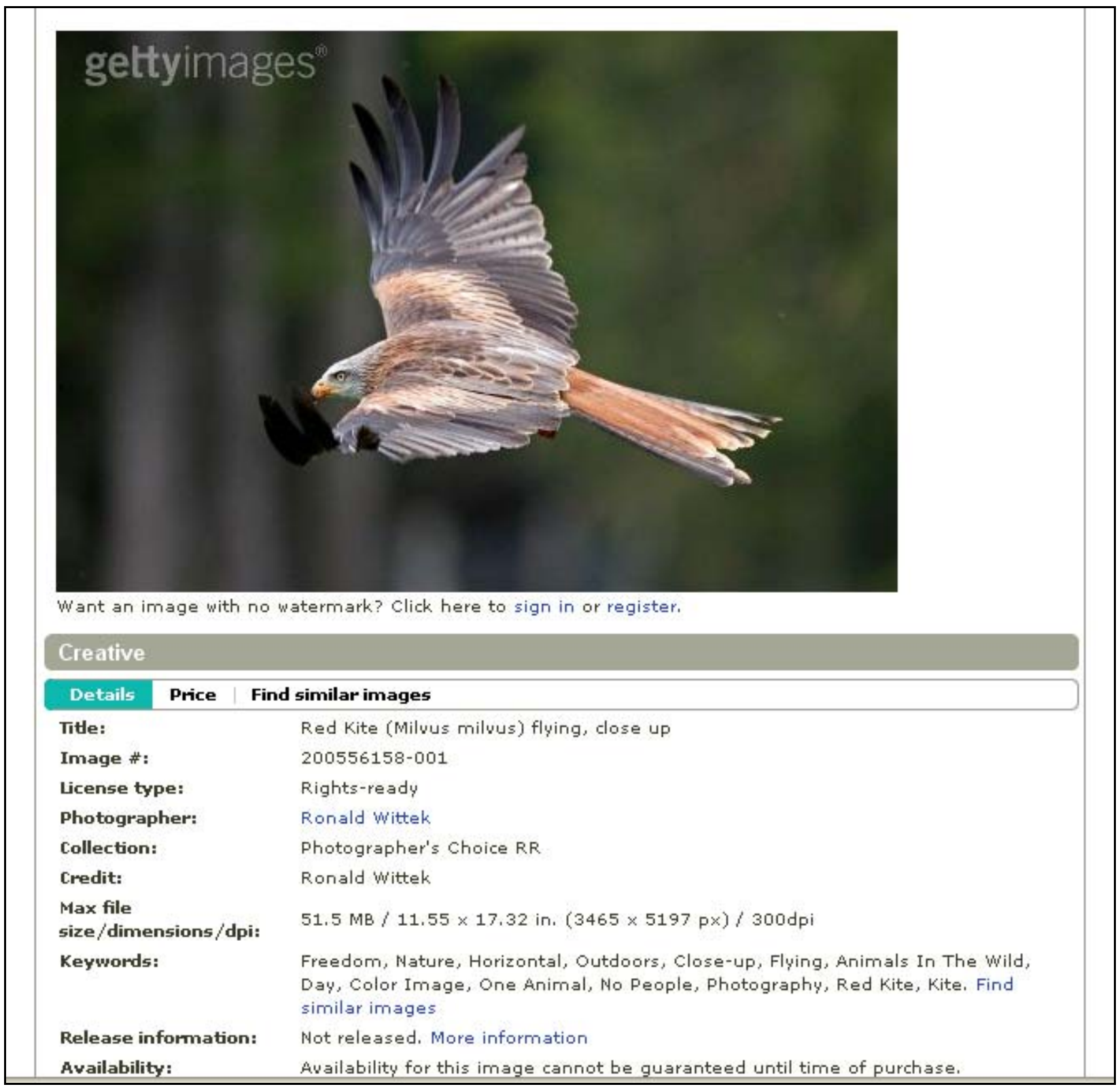